\documentclass[iop]{emulateapj}

\usepackage{pslatex}
\usepackage{amsmath}
\usepackage{textcomp}

\newcommand{\gfemail}{Gregory.A.Feiden.GR@Dartmouth.edu}
\newcommand{\ademail}{aaron.dotter@gmail.com}
\newcommand{\msun}{M_{\odot}}
\newcommand{\rsun}{R_{\odot}}
\newcommand{\teff}{T_{\rm eff}}
\newcommand{\amlt}{\alpha_{\rm MLT}}

\shortauthors{Feiden \& Dotter}
\slugcomment{Accepted to the Astrophysical Journal}

\usepackage[dvips]{hyperref}
\hypersetup {
    bookmarks=true,                    
    pdftitle={The Interior Structure Constants as an Age Diagnostic for Low-Mass, Pre-Main Sequence Detached Eclipsing Binary Stars},                  
    pdfauthor={Gregory A. Feiden &
               Aaron Dotter},          
    colorlinks=true,                   
    linkcolor=blue,                    
    citecolor=blue,                    
    urlcolor=blue                      
}
\begin{document}

\title{The Interior Structure Constants as an Age Diagnostic for Low-Mass, Pre-Main Sequence Detached Eclipsing Binary Stars}

\author{Gregory A. Feiden}
\affil{Department of Physics and Astronomy, Dartmouth College, 6127 Wilder Laboratory,\\ Hanover, NH 03755, USA; \href{mailto:\gfemail}{\gfemail}}
\author{Aaron Dotter}
\affil{Research School of Astronomy and Astrophysics, The Australian National University,\\ Weston, ACT 2611, Australia; \href{mailto:\ademail}{\ademail}}

\begin{abstract}
We propose a novel method for determining the ages of low-mass, pre-main 
sequence stellar systems using the apsidal motion of low-mass detached 
eclipsing binaries. The apsidal motion of a binary system with an eccentric 
orbit provides information regarding the interior structure constants of 
the individual stars. These constants are related to the normalized stellar 
interior density distribution and can be extracted from the predictions 
of stellar evolution models. We demonstrate that low-mass, pre-main sequence stars 
undergoing radiative core contraction display rapidly changing interior 
structure constants (greater than 5\% per 10 Myr) that, when combined with observational 
determinations of the interior structure constants (with 5 -- 10\% precision), 
allow for a robust age estimate. This age estimate, unlike those based on 
surface quantities, is largely insensitive to the surface layer where effects 
of magnetic activity are likely to be most pronounced. On the main sequence, where 
age sensitivity is minimal, the interior structure constants provide a 
valuable test of the physics used in stellar structure models of low-mass 
stars. There are currently no known systems where this technique is applicable.
Nevertheless,
the emphasis on time domain astronomy with current missions, such as $Kepler$, 
and future missions, such as LSST, has the potential to discover systems 
where the proposed method will be observationally feasible.
\end{abstract}


\section{Introduction}
In the study of the structure and evolution of low-mass stars, there are 
a variety of different methods capable of yielding reasonably accurate age 
estimates; for a thorough discussion of the different methods, and their 
strengths and weaknesses, see the review by \citet{Soderblom2010}. One
such method uses detached eclipsing binaries (DEBs) to assign an age. DEBs
are fantastic systems for studying stellar evolution.
Observations can provide precise masses and radii for the component 
stars that are nearly model independent \citep[see reviews by][]{Andersen91,
Torres2010}. Tight constraints on the stellar masses and radii allow for 
stringent tests of stellar evolution models. Furthermore, the age of a 
DEB system can be derived if stellar models can predict the radius of each 
star in the binary at a common age and with a single chemical composition.

Deriving an age estimate from a DEB is straightforward once precise masses
and radii are extracted from the data. However, the reliability of the age 
estimate is contingent upon the accuracy of the stellar models. Recently, 
pre-main sequence (pre-MS) and MS models of low-mass stars ($< 0.8 \msun$) 
have received substantial criticism for not accurately predicting the radii
of stars in DEBs. As the number of DEBs with precisely measured masses and
radii has increased, it has become clear that stellar models under-predict 
the radii of stars in DEBs by upward of 10\% \citep[see,
for example,][]{Mathieu2007,Jackson2009,Torres2010,Feiden2012}. The discrepancies
between model and observed radii have been largely attributed
to the effects of magnetic fields and magnetic activity \citep{Ribas2006,
Chabrier2007,Jackson2009,Morales2010}. When present, the discrepancy between 
observations and model predictions severely limits the use of stellar evolution 
models to derive the age of individual DEB systems.

We propose a novel method to use low-mass, pre-MS DEBs to estimate the 
ages of young stellar systems. Instead of comparing individual stellar surface
properties to stellar evolution models, we propose to use the dynamics of 
the DEB system. That is, comparing the observed rate of apsidal motion
to stellar model predictions computed using the interior structure
constants. This technique is less sensitive to the surface effects of magnetic
fields than are methods that only invoke the stellar radius or effective 
temperature. Our method has the potential to provide a more reliable age
estimate. 

Below, we outline the model calculations (Section~\ref{sec:model}), 
including the computation of the interior structure constants (Section~\ref{sec:isc}). 
Results pertaining to the time evolution of the interior structure constants 
as a function of stellar mass are given in Section~\ref{sec:res}. We 
conclude with a discussion of the usefulness of this method in Section~\ref{sec:disc}.

\section{Models}
\label{sec:model}

\subsection{Microphysics}
Model evolutionary tracks used in this study were computed with the Dartmouth 
Stellar Evolution Program (DSEP).\footnote{Available at
\url{http://stellar.dartmouth.edu/models/}}
The physics incorporated in the models have been described extensively in 
the literature \citep{Chaboyer1995,Chaboyer2001,Bjork2006,Dotter2007,
Dotter2008,Feiden2011}, but we will provide a brief overview of the physics 
pertinent to the present study.

Arguably the most critical component of low-mass stellar evolution models
is the equation of state (EOS). For masses considered in this study 
DSEP uses the FreeEOS\footnote{By 
Alan Irwin: \url{http://freeeos.sourceforge.net}} in the EOS4 configuration.  
We selected the FreeEOS for three primary reasons. First, it includes 
non-ideal contributions to the EOS, such as Coulomb interactions and 
pressure ionization, that become important in the dense plasma of low-mass 
stars. Second, the FreeEOS calculates the EOS for hydrogen, helium, and 
eighteen heavier elements as opposed to only calculating the EOS
for hydrogen and helium. Finally,
the FreeEOS may be called directly from within the stellar evolution code.
This feature avoids the need to interpolate within EOS tables, thereby
minimizing numerical errors.

Conditions near the outer, optically thin layers of low-mass stars preclude
the use of gray atmosphere approximations \citep[and references therein]{CB00}. 
Therefore, we use the {\sc phoenix ames-cond} model atmospheres 
\citep{Hauschildt1999a, Hauschildt1999b} to define the surface boundary
conditions for our interior models. The atmosphere models are attached 
at the photosphere,\footnote{The choice of where to attach the model atmosphere 
to the interior model is an important one. Our experience indicates that 
attaching the atmosphere at $T = \teff$ is reasonable for $M \ga 0.2\, \msun$; 
for lower masses it is necessary to attach the model atmosphere deeper 
into the interior.} taken to be the point where $T = \teff$. 

The radiative opacities we adopt are the OPAL opacities above $10^4$ K 
\citep{Iglesias1996} in combination with the \citet{Ferguson2005} opacities 
below $10^4$ K. For models that are not fully-convective, helium 
and heavy element diffusion are treated according to the formulation of 
\citet{Thoul1994}. Fully-convective models are assumed to be completely 
and homogeneously mixed because the convective timescale is considerably 
faster than the diffusion timescale \citep{Michaud1984}.

\subsection{Solar Calibration}
The primary input variables of stellar evolution models are defined relative to 
the Sun. These input variables include the stellar mass, the initial mass fractions of helium 
($Y_{i}$) and heavy elements ($Z_{i}$), and the convective mixing-length 
($\amlt$). Therefore, we must first define what constitutes the Sun for our 
model setup. We require that a 1 $\msun$ model accurately predict the solar radius, 
the solar luminosity, the radius to the base of the solar convection zone,
and the solar photospheric ($Z/X$) at the solar age \citep[4.57 Gyr;][]{Bahcall2005}
By iterating over different combinations of $Y_{i}$, $Z_{i}$, and $\amlt$ 
we are able to converge upon a solution for the Sun. The final set of variables 
that satisfies the above criteria for the solar heavy element composition 
of \citet{GS98} was $Y_{i} = 0.27491$, $Z_{i} = 0.01884$, and $\amlt = 1.938$.

\section{Interior Structure Constants}
\label{sec:isc}
The distribution of mass within a star in a close binary system is influenced 
by the star's rotation and by tidal interaction
with its companion. Imagine two stars, A and B. The rotation of star A
and the tidal interaction of star B with star A distorts the shape of star A.
Instead of remaining spherically symmetric, the equilibrium configuration 
of star A will become ellipsoidal. Subsequently, the gravitational potential
of star A will also become ellipsoidal. The same can be said from the 
perspective of star B.

If the binary orbit is elliptical, the distorted gravitational potential 
will cause the orbit to precess. This precession 
may be likened to the precession of Mercury's orbit about the Sun---although, 
Mercury's precession is due to general relativity. The precession
of the binary orbit is known as apsidal motion.

The rate of apsidal motion (${\dot\omega}$; measured in degrees per cycle),
or the rate at which the orbit precesses,
is governed by the shape of the gravitational potential, which may be
deformed as discussed above. Therefore, 
${\dot\omega}$ depends on the properties of the stars and of the orbit. Explicitly,
\begin{equation}
{\dot\omega} = \left(\frac{c_{2,\,1} + c_{2,\,2}}{360}\right)
                 \overline{k_{2}},
\label{eq:ap}
\end{equation}
where
\begin{equation}
c_{2,\, i} = \left[\left(\frac{\Omega_i}{\Omega_K}\right)^2 \left(
    1 + \frac{m_{3-i}}{m_i}\right) f(e) + \frac{15 m_{3-i}}{m_i} g(e)
    \right]\left(\frac{R_i}{A}\right)^5.
\label{eq:cs}
\end{equation}
In the above equation, $\Omega_K$ is the mean orbital angular velocity,
$\Omega_i$, $m_i$, and $R_i$ are the rotational velocity, the mass, and the radius 
of the $i$-th component in the binary, respectively. Additionally, $A$ is the 
semi-major axis of the orbit,
\begin{equation}
f(e) = (1 - e^2)^{-2},
\end{equation}
and 
\begin{equation}
g(e) = \frac{(8 + 12e^2 + e^4)}{8} f(e)^{5/2},
\end{equation}
with $e$ being the eccentricity of the orbit. Finally, the last term in
Equation (\ref{eq:ap}), $\overline{k_{2}}$, is the weighted
interior structure constant observed for the two binary stars. In general,
\begin{equation}
\overline{k_{2}} = \frac{c_{2,\,1}k_{2,\,1} + c_{2,\,2}k_{2,\,2}}{c_{2,\,1}
                    + c_{2,\,2}}.
\label{eq:k2w}
\end{equation}
Here, $k_{2,\,1}$ and $k_{2,\,2}$ are the interior structure constants for
each star.

Equations (\ref{eq:ap})--(\ref{eq:k2w}) are derived from a $j$-th order solid 
harmonic expansion of the gravitational potential \citep[see, for
example,][]{Kopal1978}.
The interior structure constant for a given star, $k_2$, is the second-order 
term from a more general set of expansion coefficients, $k_j$. These second-order 
coefficients quantify the central concentration---or radial distribution---of mass 
within a star \citep{Kopal1978}. Lower values of $k_2$ correspond to a 
higher level of central mass concentration. Point sources, for example, 
have $k_2 = 0$.

Observationally, we can not solve for the individual $k_{2,\, i}$ values.
However, in the above equations, every variable is a direct observable, except
for $\overline{k_{2}}$. The latter must be inferred from observational
determinations of ${\dot\omega}$ and the $c_{2,\, i}$ coefficients. Since
the individual $k_{2,\,i}$ values depend on the stellar density distribution, 
they can provide deep insight into the validity of stellar
evolution models. To bypass the restriction that only $\overline{k_{2}}$
can be inferred from observations, we use $k_2$ values from stellar 
evolution models in combination with the observed $c_{2,\, i}$ coefficients 
to derive a theoretical $\overline{k_{2}}$.

DSEP is equipped to calculate the general interior structure constants, $k_{j}$, 
for a star at every evolutionary time step. This is achieved by solving 
Radau's equation after each model iteration. Following the formalism 
outlined by \citet{Kopal1978},
\begin{equation}
a \frac{d\eta_{j}}{da} + \frac{6\rho (a)}{\left<\rho\right>}\left(\eta_{j} + 1\right) + \eta_{j}\left(\eta_{j} - 1\right) = j(j+1)
\end{equation}
with $j \in \{ 2, 3, 4, \ldots \}$ being the order of the solid harmonic,
$a$ is the radius of an equipotential surface ($a = r$ when the 
surfaces are spherically symmetric), and
\begin{equation}
\eta_{j}(a) = \frac{a}{\epsilon_{j}}\frac{d\epsilon_{j}}{da}.
\end{equation}
In the above equation, $\epsilon_{j}$ is the stellar deviation from sphericity.
We also introduced
$\rho(a)$, the density of the stellar plasma at radius $a$, and $\left<\rho\right>$, 
the volume averaged density at each radius,
\begin{equation}
\left<\rho\right> = \frac{3}{a^{3}}\int_{0}^{a}\rho (a^{\prime}) a^{\prime 2} da^{\prime}.
\end{equation}

If we assume that tides and rotation do not significantly alter the shape
of a star, we are permitted to use a spherically symmetric model to compute
the interior structure constant. 
Our code employs a $4^\mathrm{th}$-order Runge-Kutta integration 
scheme to obtain a particular solution of Radau's equation at the stellar 
photosphere. Interior structure constants are then directly related to the 
particular solutions at the surface, $\eta_{j}(R)$, through
\begin{equation}
k_{j} = \frac{j + 1 - \eta_{j}(R)}{2\left[j + \eta_{j}(R)\right]}.
\end{equation}
But, is it reasonable to assume that the stellar mass is not significantly
redistributed due to rotation and tides?

The effect of rotation on the central 
mass concentration of stars with a total mass greater than $0.8\, \msun$ 
was investigated in several previous studies \citep{Stothers1974,
cg93,claret1999}. For stars less massive than $0.8\, \msun$, the effect 
of rotation should be negligible because they have a higher mean density compared to 
solar-type stars. To test this assumption, we used Chandrasekhar's analysis of 
slowly rotating polytropes to estimate the amount of oblateness---or 
deviation from spherical symmetry---rotating low-mass stars may be expected 
to have. 

\citet{Chandra1933} derived an analytical expression for the stellar 
oblateness for slowly rotating polytropes. The oblateness was defined to 
be the relative difference between the equatorial radius and the polar
radius,
\begin{equation}
\mathcal{F} \equiv \frac{r_{\rm eq} - r_{\rm pole}}{r_{\rm eq}},
\end{equation}
with $r_{\rm eq}$ and $r_{\rm pole}$ being the equatorial and polar radius, 
respectively. Polytropes were considered slowly rotating when 
\begin{equation}
\chi \equiv \frac{\Omega^2}{2\pi G \rho_c} \ll 1,
\end{equation}
where $\Omega$ is the stellar angular velocity, $G$ is the gravitational 
constant,  and $\rho_c$ is the mass density in the stellar core. Does this
criterion apply to real low-mass stars? If we assume a rotation period 
of 1.0 day and a core density of 10 g~cm$^{-3}$ (realistic for pre-MS low-mass
stars), then $\chi \approx 10^{-3}$. 

The result of Chandrasekhar's analysis was a relation between the stellar
oblateness and $\chi$ for different values of the polytropic index, $n$,
\begin{equation}
\mathcal{F} = \left\{
    \begin{array}{r c l}
        5.79\chi  & & {\rm for}\, n = 1.5 \\
        9.82\chi  & & {\rm for}\, n = 2.0 \\
        41.8\chi  & & {\rm for}\, n = 3.0 \\
    \end{array}
    \right. .
\end{equation}
Stars with masses below 0.65~$M_{\odot}$ are best represented by a polytrope
with $1.5 < n < 2.0$. Assuming $n = 2.0$, a low-mass star will have 
$\mathcal{F} \sim 0.01$. This treatment indicates that the effect of rotation 
on the sphericity of low-mass stars is a 1\% effect; rotation will not be addressed 
in this study.

Assessing the influence of tides is difficult. We expect spherically
symmetric models to provide accurate estimates of the mass distribution if
$R_{*} \ll R_{\rm roche}$, where $R_{\rm roche}$ is the Roche lobe 
radius.

We also caution that the configuration of the binary must be considered.
If the rotational axes are not aligned, as found with \object{DI Her} \citep{Albrecht2009}, 
then the validity of the assumptions required to derive of $k_2$ no longer hold.
See Section \ref{sec:disc} for a further discussion.

General 
relativistic distortion of the gravitational potential also plays a role 
in determining the rate of apsidal motion \citep{gimen1985}. This contribution 
to the apsidal motion rate can be added to the classical contribution (i.e., 
Equation~(\ref{eq:ap}))
\begin{equation}
{\dot\omega}_{\rm tot} = {\dot\omega}_{N} + {\dot\omega}_{\rm GR},
\end{equation} 
where ${\dot\omega}_{\rm tot}$, ${\dot\omega}_{N}$, and ${\dot\omega}_{\rm GR}$
are the total apsidal motion rate, classical apsidal motion rate, and the rate
predicted from general relativity. The general relativistic contribution does 
not depend on the mass distribution of the stars. Thus, general relativity 
does not affect the theoretical derivation of the interior structure constants,
but must be accounted for prior to comparing the theoretical and observational
determinations of $\overline{k_{2}}$.

\begin{figure*}
    \plotone{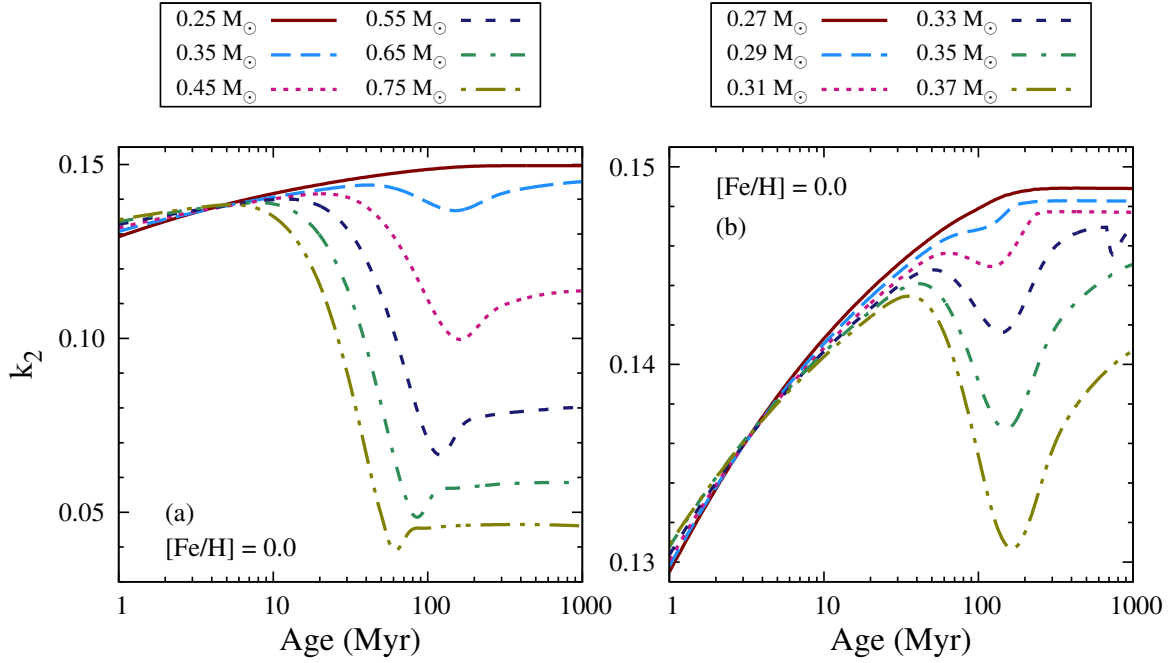}
    \caption{The time evolution of the interior structure constant, $k_2$,  
    for stars of various masses. Stars that develop a radiative core during 
    the pre-main sequence exhibit rapidly decreasing $k_2$ values between roughly 10 
    and 100 Myr. (a) Total range of masses considered in this 
    study shown in increments of 0.05~$M_{\odot}$. (b) A detailed 
    view of the transition to fully-convective interiors. Only the 
    0.37~$M_{\odot}$ model does not ultimately become fully-convective.}
    \label{fig:k2age}
\end{figure*}

\begin{figure*}
    \plotone{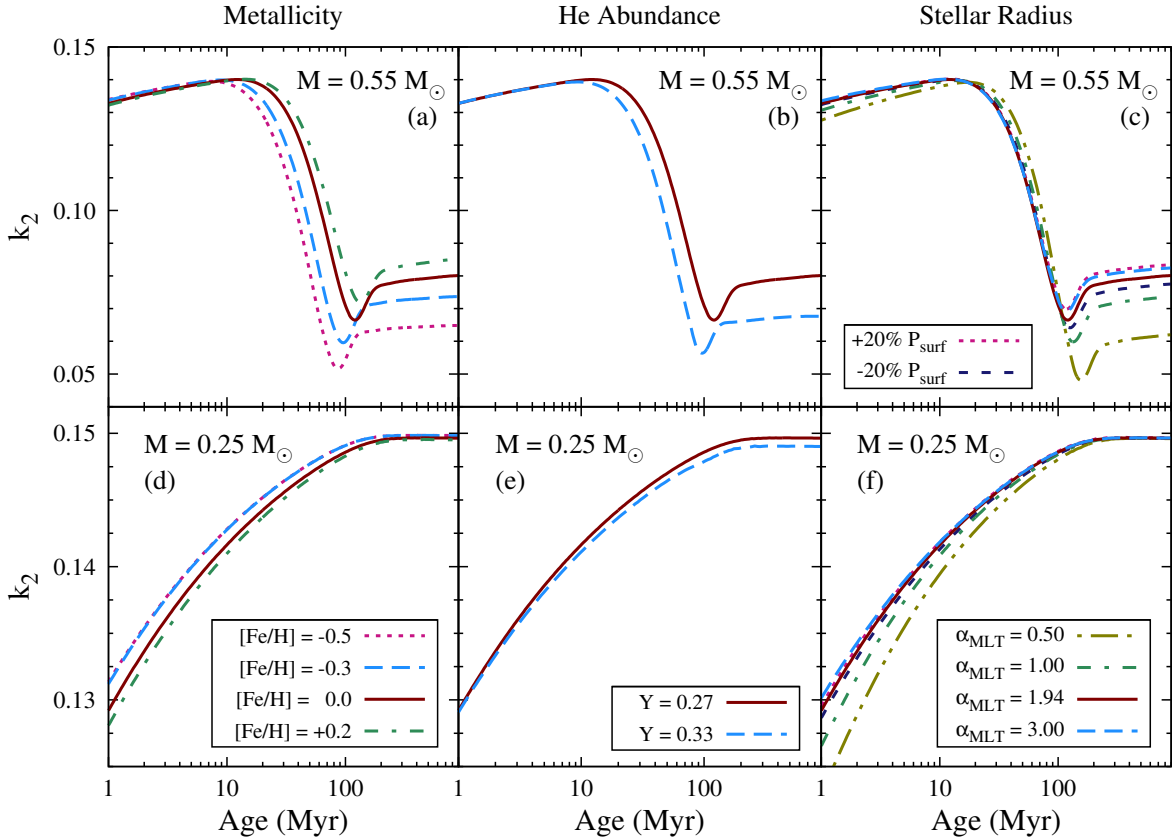}
    \caption{The influence of various model properties on the predicted 
    evolution of $k_2$ for two different stellar masses. From left to right, 
    the properties investigated are: scaled-solar metallicity in (a) and 
    (d), helium mass fraction ($Y$) in (b) and (e), and artificial radius 
    changes in (c) and (f). Models in the top panels ((a)--(c)) show a 
    0.55~$M_{\odot}$ star while the bottom series ((d)--(f)) show a 
    fully-convective 0.25~$M_{\odot}$ star. Note, the legend for panels 
    (c) and (f) is split between the top and bottom panels due to space 
    restrictions. The separate tracks indicated by both legends 
    are presented for each mass.}
    \label{fig:k2prop}
\end{figure*}

\section{Results}
\label{sec:res}

\subsection{Single Stars}
\label{sec:res_1}
Individual, solar-metallicity evolutionary tracks were computed for several 
masses ranging from the fully-convective regime ($0.25~M_{\odot}$) up to 
masses where stars have thin convective envelopes ($0.75~M_{\odot}$). Near 
the boundary where the transition from a radiative core to a fully-convective 
interior is expected ($\sim0.35~M_{\odot}$), a finer grid of mass tracks 
was generated to allow for further exploration. The evolution of $k_2$ with 
age for each mass track is presented in Figure~\ref{fig:k2age}. The full 
collection of mass tracks (including analysis routines) used hereafter have been 
made available on online\footnote{
\url{https://github.com/gfeiden/k2age/}}.

Stellar models that develop a radiative core have a rapidly-changing interior 
structure constant between the age of 10 and 100~Myr. This can be observed 
in Figure~\ref{fig:k2age}(a). As the convection zone recedes, 
the central regions of the collapsing pre-MS star create a more centrally-concentrated 
mass profile, lowering the derived value of $k_2$. The result is that 
the value of $k_2$ decreases by about 5\% -- 10\% every 10~Myr for 
masses above 0.45~$M_{\odot}$. Masses below approximately 0.45~$M_{\odot}$ 
undergo variations up to about 5\%. 

This period of rapid contraction continues until a small convective core develops, 
producing a star with three energy transfer zones: a convective core, a 
radiative shell, and a convective outer envelope. The star settles onto 
the MS once the small convective core subsides with the 
equilibration of $^3$He burning. In Figure~\ref{fig:k2age}, this process
is manifested by the upward turn of the mass tracks near 100~Myr, 
followed by a flattening of $k_2$ as the star enters the MS. 

After first developing a small radiative core on the pre-MS, stars with 
masses between 0.29~$M_{\odot}$ and 0.35~$M_{\odot}$ eventually maintain 
a fully-convective interior (see Figure~\ref{fig:k2age}(b)). 
These small radiative cores manifest themselves as small dips in the evolution 
of $k_2$, just as for stars that maintain a radiative core on the MS. While 
the rate of change in $k_2$ is rapid during the core contraction, the relative 
change in $k_2$ with stellar age is small ($<~5\%$). It is evident from 
Figure~\ref{fig:k2age}(b) that near the fully-convective transition, stars 
that eventually end up with fully-convective interiors exhibit a degeneracy 
in the age-$k_2$ plane. There is no differentiating between their pre-MS core 
contraction and the eventual reduction of the radiative shell using $k_2$ 
alone. Below 0.29~$M_{\odot}$, a radiative core does not develop on the pre-MS.

Additional models were generated to investigate the effect of specific 
stellar properties on the predicted values of $k_2$. 
In Figures~\ref{fig:k2prop}(a)--(f)
we illustrate the results of changing the scaled-solar stellar metallicity, 
helium abundance ($Y$), and the effects of artificially inflating and deflating 
the stellar radius. For this exercise, two masses were selected  
to study the effects on the two broad categories of low-mass stars: 
fully-convective stars and stars with radiative cores.

It is apparent from Figure~\ref{fig:k2prop} that changes to the 
chemical composition have the largest effect on the value of the interior 
structure constant on the pre-MS. Variations in metallicity of 0.2~dex 
translate into a 5\% difference in the calculated $k_2$ for a 0.55 $M_{\odot}$ 
star at a given pre-MS age. Similarly, large variations in the helium 
abundance have the ability to produce changes in $k_2$ at the 5\% level. 

Without some prior knowledge of the stellar composition, the effects of 
such variations may be confused with a difference in stellar age. The 
effects on  $k_2$ are lessened as mass decreases, until variations nearly 
vanish in the fully-convective regime 
(panels (d) and (e) of Figure~\ref{fig:k2prop}). 
Note that a change in $Y$ pushes $k_2$ in the same direction in both 
mass regimes, whereas a change in metallicity produces a change in one 
direction for the radiative core case but the opposite direction in the 
fully-convective case.  While this behavior does not lift the age-composition 
degeneracy entirely, it could prove a useful diagnostic in DEB systems whose 
components straddle the fully-convective boundary.
 
We attempted to mimic the possible effects of magnetic fields on the structure
of our models by computing models at solar-metallicity with $\amlt$
= 0.5, 1, and 3. The modified mixing-length represents magnetic suppression
of convection in the deep interior. Additionally, models were run where 
we artificially changed the surface pressure by $\pm20$\%. The altered 
surface pressure represents the possible influence of star spots on the 
stellar photosphere. The magnetic Dartmouth models \citep{FC12b} were not
used as they have yet to be evaluated for stars on the pre-MS.

Figures~\ref{fig:k2prop}(c) and \ref{fig:k2prop}(f) show how $k_2$ varies 
with  changes to $\amlt$ and the surface pressure. During the pre-MS 
contraction, a 20\% change in the surface pressure results in a $\sim2$\% 
change to the stellar radius. The accompanying change in $k_2$ was found to be 
0.2\% and 1.0\%, for a positive and negative change to the surface pressure, 
respectively. In the $0.55~M_{\odot}$ models, increasing $\amlt$ 
to 3 yielded a stellar radius that was 2\% smaller than our solar-calibrated 
model with $k_2$ variations under 1\% throughout the star's pre-MS 
contraction. Decreasing $\amlt$, however, produced larger variations 
in $k_2$. At an age of 60~Myr the model radii appeared inflated by 5\% and 
15\% for $\amlt$ = 1.0 and 0.5, respectively. The corresponding 
changes in $k_2$ were, respectively, 4\% and 12\%. Panel (f) 
of Figure~\ref{fig:k2prop} indicates that the $0.25~M_{\odot}$, fully-convective 
models experience the greatest differences at the youngest age and that 
these differences diminish until the different tracks converge on the MS.
These changes to $\amlt$ and surface pressure are for illustrative 
purposes only. It is not at all clear that simply reducing $\amlt$ 
is a suitable approximation for the presence of an interior magnetic field nor 
that altering the surface pressure is a good approximation for star spots 
at the surface.

\subsection{Binary Systems}
\label{sec:binary}

\begin{figure}
    \plotone{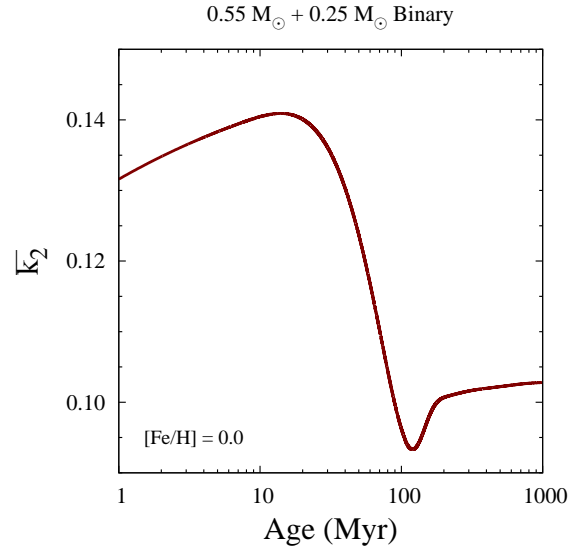}
    \caption{The evolution of the theoretical $\overline{k_{2}}$ for a binary having a
    $0.55\msun$ primary and a $0.25\msun$ secondary. The ``observed'' stellar radii are
    fixed at $0.62\rsun$ and $0.41 \rsun$ for the primary and secondary, respectively 
    (see Section \ref{sec:binary}). The orbit
    was chosen to have an eccentricity $e=0.2$.}
    \label{fig:binary}
\end{figure}

The rapid evolution of $k_2$ for a single star can provide an accurate age
estimate for that star, but what about for a binary system? We stated in
Section \ref{sec:isc} that observations are only able to provide the weighted 
mean value of $k_2$ for two stars in a binary. To derive an age estimate
for a binary system, we must find the theoretical $\overline{k_2}$. 

Temporal evolution of the theoretical $\overline{k_2}$ values can be obtained 
by combining two $k_2$ mass tracks using Equation (\ref{eq:k2w}). 
Computation of the $c_{2,\, i}$ coefficients requires precise knowledge
of the stellar masses, radii, and the orbital
eccentricity (see Equation (\ref{eq:cs}) and note that $A$ becomes irrelevant). 
The angular velocity term 
in Equation (\ref{eq:cs}) can either be measured or approximated using
the orbital eccentricity, assuming
pseudo-synchronization \citep{Kopal1978},
\begin{equation}
\Omega_i^2 = \frac{(1+e)}{\left(1-e\right)^3} \Omega_K^2 .
\end{equation}
Hence our reason for focusing on DEBs: they can yield precise estimates
of the stellar and orbital properties.

The evolution of $\overline{k_2}$ is simplest for an equal mass binary. 
In this case, both stars contribute equally to $\overline{k_2}$, meaning
the $\overline{k_2}$ track is exactly equal to the two individual $k_2$ 
tracks. The discussion from Section~\ref{sec:res_1} on single star tracks
would then apply to the binary system.

A binary with unequal mass components is not so simple. Two mass tracks
are required---one computed for the mass of each star---and must be combined as a
single track using Equation~(\ref{eq:k2w}). How does this effect our ability
to extract an age estimate?

We have provided an example of a weighted $k_2$ evolutionary track in 
Figure~\ref{fig:binary}. The masses selected are those of the two stars
presented in Figure~\ref{fig:k2prop}, $M_1 =0.55 \msun$ and $M_2 = 0.25\msun$.
For this example, we arbitrarily adopted an orbit with an eccentricity $e=0.2$. 
Weighting of the theoretical $k_2$ values is insensitive to the semi-major
axis. The stellar radii ($R_1 = 0.62\rsun$
and $R_2 = 0.41 \rsun$) were selected from a solar metallicity mass track 
at an age of 40 Myr.  To simulate the type of data available to an observer,
we elected to fix the radius, and thus, fix the $c_{2,\,i}$ coefficients
(used to weight the value of $k_2$) at each age in Figure~\ref{fig:binary}.

The rapid evolution of $k_2$ that is taking place in the $0.55\msun$ 
star largely dominates the relatively slow $k_2$ evolution of the $0.25\msun$ 
star. While the weighted $k_2$ value for the binary does not evolve as 
rapidly as for a single $0.55\msun$ star, the weighted value still changes 
by about 5\% every 10 Myr. This is comparable to models of single stars 
and should not significantly hinder any age analysis.

\section{Discussion}
\label{sec:disc}

DEBs provide an excellent laboratory for testing stellar structure and evolution 
theory in different mass and evolutionary regimes. The mass-radius plane 
is the strictest test of stellar models because these two quantities are 
best constrained by the observations \citep{Torres2010}. Results from studies 
performing such comparisons have led to the consensus that standard stellar 
evolution models are unable to accurately reproduce the observed stellar 
radii for masses below $\sim0.8 \msun$ \citep[e.g.,][]{Torres2010,Feiden2012}. 
The model inaccuracies are particularly evident among pre-MS binaries 
\citep{Mathieu2007,Jackson2009}.

Radius discrepancies of approximately 10 -- 15\% are routinely quoted between 
pre-MS DEBs and models. This makes it difficult to derive an age with less 
than 50\% uncertainty from stellar positions in the mass-radius plane. Magnetic 
effects, particularly surface spots, are thought to belie the observed radius 
deviations. Canonical stellar evolution models are non-magnetic and are 
therefore unable to properly account for magnetic modifications to convective 
energy transport and for the presence of magnetic spots on the stellar 
photosphere. 

We therefore advocate the inclusion of the interior structure constant, 
$\overline{k_2}$, to overcome these age determination inaccuracies whenever possible. 
While this study does not lead to us to conclude that individual $k_2$ values
(and thus $\overline{k_2}$) are entirely 
insensitive to magnetic activity in low-mass stars, it is evident that $\overline{k_2}$ 
has the potential to be a better diagnostic of the pre-MS evolutionary state 
than the stellar radius when surface magnetic activity is present. 

For
instance, setting $\amlt~=~0.5$ increases the radius and $k_2$ of a single star by 
15\% and 12\%, respectively. Fixing the radius to determine an age leads 
to an age that is upward of 180\% greater than if we assume a solar calibrated 
$\amlt$. On the other hand, fixing $k_2$ leads to only a 20\% greater 
age. The age errors are then compounded when we consider both stars in the 
binary. The decreased sensitivity of the individual $k_2$ values makes the
mean $\overline{k_2}$ a superior choice compared to surface quantities like
the radius and effective temperature.

The age precision returned from observational determinations of $\overline{k_2}$ is 
dependent on the precision with which the observational $\overline{k_2}$ and the system's metallicity 
are known. For example, given an equal mass binary with 0.5~$M_{\sun}$ stars, 
knowing $\overline{k_2}$ with 
5\% uncertainty and the metallicity to $\pm0.2$~dex yields a pre-MS age 
with an uncertainty of approximately 33\%. Constraining the metallicity 
uncertainty to $\pm$0.1~dex improves this age uncertainty to 20\%. Furthermore, 
to obtain an age with 5\% precision would require $\overline{k_2}$ to be measured with 
near 1\% precision for a metallicity known to within 0.1~dex.

\subsection{Observational Considerations}

The results presented and discussed above show that it is possible
to precisely derive the age of a binary system from measurements of
apsidal motion. However, obtaining precise observations of apsidal motion
and measuring $\overline{k_2}$ with 5\% precision is a painstaking task.
The binary system must meet several criteria and the data must be of
high quality. 

Foremost is that the binary \emph{must} be a double-lined,
eclipsing system. While this has been alluded to, we have yet to illuminate 
precisely why this is so. The reason for requiring a DEB stems from the need for extremely
accurate and precise stellar and orbital properties. Equations
(\ref{eq:ap}) and (\ref{eq:cs}) reveal that derivation of $\overline{k_2}$ 
requires exquisite knowledge of the stellar mass ratio, the stellar radii, and the 
orbital properties (eccentricity and semi-major axis). 
Only data from DEBs can provide these quantities in a (nearly) model-independent 
fashion \citep{Andersen91,Torres2010}. 

A meticulous examination of the binary light curve and radial velocity curve 
is of the utmost importance. The light curve must be assembled from multi-epoch,
time-series differential photometry and must provide nearly complete phase 
coverage. This latter feature is essential. Not only must the primary and 
secondary eclipses be captured, but also the behavior of the light curve
out of eclipse. Such data is becoming increasingly available with the latest
generation of photometric surveys (e.g., \emph{Kepler}, CoRoT\footnote{Convection,
Rotation, \& planetary Transits}, SuperWASP\footnote{Wide Angle Search for 
Planets}) and those to come (e.g., LSST\footnote{Large Synoptic Survey Telescope}, 
BRITE\footnote{Bright Target Explorer} Constellation Mission).

Similar to the photometry, a large number of high dispersion, high 
signal-to-noise spectra are needed to construct a detailed radial velocity
curve. Attempts must be made to provide adequate phase coverage \citep[see][]{
Andersen91} and uncover deviations due to the Rossiter-McLaughlin effect
\citep{Rossiter1924,McLaughlin1924} 
for reasons we shall discuss momentarily. \citet{Torres2013} points out that
the spectroscopy is still the limiting factor of quality in DEB analyses.

Only with the quality of data described above and within \citet{Andersen91}
and \citet{Torres2010} can a truly adequate analysis of a DEB be performed.
However, acquisition of data of such quality is very rewarding and allows 
for a rigorous examination and characterization of the stellar system. It 
would permit the measurement of the stellar masses and radii with extreme
precision (below 2\%). This is imperative considering that Equation (\ref{eq:cs}) 
depends upon the fractional radii $(R/A)$ to the 5$^{\rm th}$ power. Additionally,
actual measurement of $\dot{\omega}$ requires careful monitoring of eclipse 
times of minimum. This can only be performed if one has a densely populated 
light curve.

The data would also permit measurement of the binary eccentricity and semi-major
axis with high precision. These properties affect both Equation (\ref{eq:cs}) 
and the general relativistic contribution discussed in Section~\ref{sec:isc}
\citep{gimen1985}. Recall that the contribution from the latter must be removed
from the total apsidal motion rate to derive the classical contribution given
in Equation (\ref{eq:ap}). Detailed radial velocity curves not only provide
accurate mass and eccentricity estimates, but may also be used to investigate
the inclination of the system. The theory presented in Section \ref{sec:isc}
relies on the assumption that the rotational axes of the two stars be parallel
to one another and perpendicular to the orbital plane. It is possible to evaluate
this restriction by using the Rossiter-McLaughlin effect in a manner similar to 
that presented for DI Her \citep{Albrecht2009}. A detailed radial velocity 
curve may, additionally, betray the presence of a third body. The presence
of a tertiary may affect the binary orbit, altering the derived apsidal
motion.

One added benefit is that lengthy observations 
may help to identify---and thus correct for---the impact of star spots 
and magnetic activity on the eclipse profiles. Star spots have the ability to distort the 
light curve, which can diminish the accuracy of the derived stellar 
properties \citep{Windmiller2010}. Removing the effects of spots is critical 
to obtain not only precise but also accurate stellar properties. The easiest 
means of obtaining detailed
time-series photometry is through space-based satellites, such as CoRoT 
and \emph{Kepler}. However, long-term ground-based observational efforts 
are beginning to produce apsidal motion detections \citep[e.g.,][]{Zasche2012}, 
demonstrating that it is feasible to carry out the necessary observations
using ground-based telescopes.

Finally, by acquiring a large number of quality spectra, it may be possible to
extract the projected rotational velocities ($v\sin i$) and a modest estimate
of the chemical composition. Spectroscopic determinations of cool star 
metallicities is notoriously complicated, but most pre-MS DEBs will likely 
reside near or in a cluster from which metallicity estimates can be extracted
using the higher mass stellar population. 
In the event there is no known association from which to draw
a metallicity, techniques based on low- and medium-resolution spectra are 
encouraging \citep[see, e.g.,][]{RojasAyala2012}. While the validity 
of such techniques along the pre-MS is unclear, they provide a viable 
starting point and typically produce metallicities with uncertainties 
below $0.2$ dex, the limit we recommend. 

\subsection{Limitations}

The usefulness of $\overline{k_2}$ as an age estimator is limited to the pre-MS, in 
particular, during the evolutionary phase where the radiative core is rapidly 
contracting. This typically corresponds to an age between 10~Myr and 
100~Myr (see Figure~\ref{fig:k2age}). At the 5 -- 10 \% measurement level, 
this technique is also restricted to binaries where one of the stars has
a mass above $\sim$0.40~$M_{\odot}$. This restriction ensures
the radiative core contraction and relative change in $k_2$ for the more 
massive star is rapid 
enough to dominate the theoretical $\overline{k_2}$ evolution. Reducing the observational uncertainty 
in $\overline{k_2}$ below 5\% enables a more 
accurate age estimate and would allow for DEBs with lower mass components 
to be reliably analyzed. 


Precise measures of apsidal motion and metallicity are challenging to obtain, but
are already feasible and should only improve over time. Instead of observational limitations, 
the greatest limiting factor for using $\overline{k_2}$ as an age indicator 
is the circularization of the DEB orbit. Mutual tidal interactions will 
circularize binary orbits over time \citep{zahn1977}. Apsidal motion
requires an elliptical orbit. If an equal-mass binary is to maintain an 
elliptical orbit for the duration of its pre-MS contraction, the orbital
period must be at least 2.4 days \citep{zahn1977}. The probability of 
discovering an elliptical binary decreases with time, but this provides 
an additional consistency check. The age suggested by $\overline{k_2}$ 
should not be significantly 
older than the orbital circularization timescale.

Finally, the age estimate is only as accurate as the stellar models. Verifying 
that a given stellar model produces the proper mass distribution, hence $k_2$,
may at first seem rather unreasonable. However, at least one known system, with 
the possibility of a second \citep[\object{KIC 002856960};][]{Lee2012}, is 
capable of providing validation of the physics incorporated in low-mass 
stellar evolution models. \citet{Carter2011} have indicated that by the 
end of the nominal \emph{Kepler} mission, they will know the interior 
structure constants of \object{KOI-126} B and C with about 1\% precision. Interior 
structure constants known with this precision can place stringent constraints
on the equation of state of the stellar plasma \citep{Feiden2011}. The 
veracity of low-mass models, and therefore the validity of their predicted 
interior structure constants, may be assessed according to the results 
from KOI-126 and similar systems.

\acknowledgements
The authors thanks G.\ Torres and K.\ Stassun for insightful conversations,
the anonymous referee for helpful comments and suggestions,
and Alan Irwin for his work on the open source FreeEOS project. G.A.F.\ 
also thanks the Department of Physics and Astronomy at Uppsala University
for their gracious hospitality during the completion of the manuscript.
G.A.F.\ acknowledges the support of the William H. 
Neukom 1964 Institute for Computational Science and the the National 
Science Foundation (NSF) grant AST-0908345. 
A.D.\ received support from the Australian Research Council under grant FL110100012.
This research has made use of NASA's Astrophysics Data System.

\end{document}